\documentclass[twocolumn,amssymb,showpacs, amsmath,nobibnotes,nofootinbib,aps,prb]{revtex4}
\usepackage[dvips]{graphicx}
\begin{document}
\title{Magnetic susceptibility and heat capacity investigations of the unconventional spin-chain compound Sr$_3$CuPtO$_6$}
\author{S. Majumdar}
\author{V. Hardy}
\altaffiliation[Permanent address: ]{Laboratoire CRISMAT, UMR 6508, Boulevard du Mar\'echal Juin, 14050 Caen cedex, France.}
\author{M. R. Lees} 
\author{D. M$^c$K. Paul}
\affiliation{Department of Physics, University of Warwick, Coventry CV4 7AL, United Kingdom}
\author{H. Rousseli\`ere}
\author{D. Grebille}
\affiliation{Laboratoire CRISMAT, UMR 6508, Boulevard du Mar\'echal Juin, 14050 Caen cedex, France}
\begin{abstract}
The Heisenberg spin chain compound  Sr$_3$CuPtO$_6$ is investigated by magnetic susceptibility and heat capacity measurements. Sr$_3$CuPtO$_6$ has an unconventional chain structure in the sense that i) the  spin half copper atoms are arranged in a zigzag chain structure, and ii)  neighboring Cu atoms along the chains  are separated by  spin zero platinum atoms. We report that this compound shows broad  features in the temperature dependence of  both the magnetic contribution to the  heat capacity and the magnetic susceptibility. Despite the unconventional nature of the spin chain structure, this set of data exhibits good agreement with theoretical models for a classical  $S = \frac{1}{2}$ Heisenberg spin chain compound.  The values of the intra-chain coupling constant, obtained by different techniques, are found to be very close to each other. The low temperature heat capacity data (below $\sim$ 6 K) exhibit a deviation from the theoretically expected behavior, which could be related to a small energy gap in the spin excitation spectrum. 
\end{abstract}

\maketitle

\section{Introduction}
A new class of one-dimensional (1D) magnetic compounds with the  general formula A$_3$XYO$_6$ (A = Sr, Ca; X, Y = transition metals) has attracted particular interest recently.~\cite{stitzer} These materials crystallize in a K$_4$CdCl$_6$-type  rhombohedral structure, which consists of infinite chains built up of alternating face-sharing YO$_6$ octahedra and XO$_6$ trigonal prisms. These  YO$_6$-XO$_6$ chains are separated from each other by A$^{2+}$ cations and  constitute a hexagonal cross-sectional  arrangement perpendicular to the chain direction. Depending upon the nature  and the oxidization states  of the  cations at the X and the Y sites, these materials present a  range of low dimensional magnetic phenomena.~\cite{nguyen} In this class of materials,  Sr$_3$CuPtO$_6$ is reported to be  a  spin half  Heisenberg antiferromagnet (AFM) chain compound.~\cite{claridge,nguyen,science} 

\par
From a crystallographic point of view, Sr$_3$CuPtO$_6$ does not  constitute an ideal linear spin chain structure~\cite{hodeau}. Firstly, due to the presence of  Cu$^{2+}$ cations at the  X site, the K$_4$CdCl$_6$-type rhombohedral structure undergoes a monoclinic distortion. The Pt$^{4+}$ ions stay at the center of the PtO$_6$ octahedra maintaining a linear structure, whereas the Cu$^{2+}$ ions move from the center of the CuO$_6$ trigonal prisms to their faces. Successive  Cu$^{2+}$ ions on a chain are displaced away from  the chain axis in opposite directions, resulting in a zigzag chain structure with a Cu-Cu bond angle of 160.94$^{o}$ (see Fig. 1). Secondly, along each chain, neighboring $S = \frac{1}{2}$ Cu$^{2+}$ ions are separated by a  Pt$^{4+}$ ion. In the presence of a strong octahedral crystal field, the Pt$^{4+}$ ions adopt a low spin ground state with zero effective spin. This results in  a   $\frac{1}{2}$---0---$\frac{1}{2}$---0---$ \frac{1}{2}$..... type of spin chain structure in which  the $S = \frac{1}{2}$ inter-spin separations parallel and perpendicular to the chains are almost equal. Prompted by this fact,  Claridge {\it et al.}~\cite{claridge} analyzed the magnetic susceptibility data of Sr$_3$CuPtO$_6$  considering both intra-chain ($J$) and inter-chain ($J'$) couplings among the Cu$^{2+}$ ions. From their analysis, using  a modified Bonner and Fisher model~\cite{hatfield}, they reported  that $J$ = 24.7 K and $J'$ = 7.3 K. Since $J$/$J'$ $\sim$ 3, they  concluded that the compound  should not be considered as a 1D magnetic system. However, the broad  maximum observed  in the magnetic susceptibility data~\cite{claridge, nguyen} gives a strong indication that the compound does indeed have the characteristics of a low-dimensional magnetic system. As a result, the magnetic status of Sr$_3$CuPtO$_6$ remains   unclear. 
\begin{figure}[t]
\centering
\includegraphics[width = 6 cm]{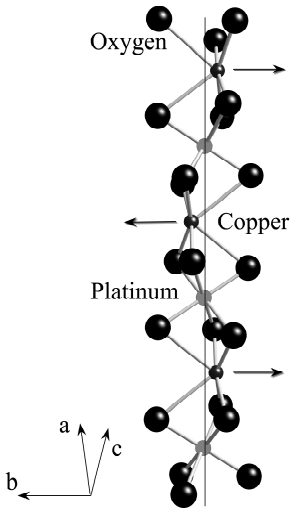}
\caption{A single CuO$_6$-PtO$_6$ chain in the crystal structure of  Sr$_3$CuPtO$_6$. The line through the platinum atoms indicates the direction of the chain axis. The arrows indicate the displacement of the copper atoms from the chain axis resulting in the formation of a zigzag chain structure. Each chain is separated from  the others by the Sr$^{2+}$ ions.}
\label{fig1}
\end{figure}

\par
In this paper, we present the results of a thorough investigation of the magnetic susceptibility and heat capacity behavior of single crystal samples of Sr$_3$CuPtO$_6$. While there have been  studies of  the magnetic susceptibility of this compound~\cite{claridge, nguyen, irons}, the present paper reports the first investigation of the heat capacity of this material. Despite the fact that heat capacity  can provide important information about the magnetic character of spin chain systems, very little heat capacity data have been published to date for the A$_3$XYO$_6$ class of compounds~\cite{niazi,hardy, rayprol,kausik}.

\par
The theoretical model of Bonner and Fisher has been successfully used in the past to fit the experimental  susceptibility data for $S = \frac{1}{2}$ spin chain systems~\cite{bonner}. This model is based on  the extrapolation of  numerical calculations for a finite number of spins ($\leq$ 11) and is valid in a limited temperature range ($T \geq \frac{1}{2}J$, where $J$ is the intra-chain coupling constant in kelvin). Recently Kl\"umper, Johnston, and co-workers~\cite{johnston,klumper} have used the Bethe anstatz approach to formulate  a model for the $S = \frac{1}{2}$ Heisenberg spin chain system, which should be  valid over the entire temperature range. This model is an  extension of previous theoretical work~\cite{eat,lukyanov}. In the present paper, the majority of the fitting has been carried out using this new  analytical model proposed by Johnston and co-workers (hereafter called the Johnston model)~\cite{johnston}.  The Bonner-Fisher model has  also been used  for comparison. Throughout the paper, we have adopted the convention where  the spin Hamiltonian, $\mathcal{H}$, is defined as $\mathcal{H}$ = $2Jk_B\sum_{i} {\bf S}_i. {\bf S}_{i+1}$ , where ${\bf S}_i$ is  the $i$-th  spin on the chain and $k_B$ is the Boltzmann constant. 

\section{Experimental details}
The needle shaped crystals of Sr$_3$CuPtO$_6$ used in this investigation were  prepared by a K$_2$CO$_3$ flux  method as described in reference~\cite{claridge}. The crystals were analyzed by  x-ray diffraction using Mo $K_\alpha$ radiation on a Bruker Nonius Kappa CCD single crystal x-ray diffractometer. The cell refinement showed that the crystals have a monoclinic structure (space group C2/c) with lattice parameters $a$ = 9.34(1) \AA, $b$ = 9.64(1) \AA, $c$ = 6.68(1) \AA~  and $\beta$ = 91.91(5)$^o$, which are in good agreement with the literature~\cite{hodeau}. The rod axis of the crystals was found to be along the [201] direction, which coincides with the CuO$_6$-PtO$_6$ chain direction. The magnetic susceptibility  measurements were carried out using  a Quantum Design SQUID magnetometer in the temperature ($T$) range 2-400 K with the applied magnetic field ($H$)  parallel  to the chain direction. The zero field heat capacity ($C$)  measurements were performed  by a relaxation method using a Quantum Design Physical Properties Measurement System (PPMS). The $C$ versus $T$ data in the temperature range  1.9 to 300 K were recorded with a conventional $^4$He insert, whereas, the data collected  below 1.9 K were recorded using a $^3$He refrigerator option. Since the mass of each crystal is  typically $\sim$ 1-2 mg, an assembly of 6 crystals, with their rod axis aligned, were used for both the magnetization ($M$) and the heat capacity measurements.

\par
In order to obtain information about the short range magnetic correlations from the heat capacity data, it is necessary  to estimate  the lattice contribution to the heat capacity ($C_{lat}$). It is therefore, important to measure a non-magnetic reference compound with the  same crystal structure and containing elements with  similar masses. For this purpose, a ceramic sample of  the isostructural compound  Sr$_3$ZnPtO$_6$ was  prepared by the usual solid state reaction technique~\cite{christina}. This compound was found to be essentially nonmagnetic. The small Curie constant $c_{cw} \sim 2.6 \times 10^{-3}$ emu/K-mol, derived from the  magnetic susceptibility versus temperature data for Sr$_3$ZnPtO$_6$, corresponds to only  0.1\% of the Pt$ ^{4+}$  ions on the  prismatic sites being in the high-spin ($S$ = 2) magnetic state.

\begin{figure}[t]
\centering
\includegraphics[width = 7.5 cm]{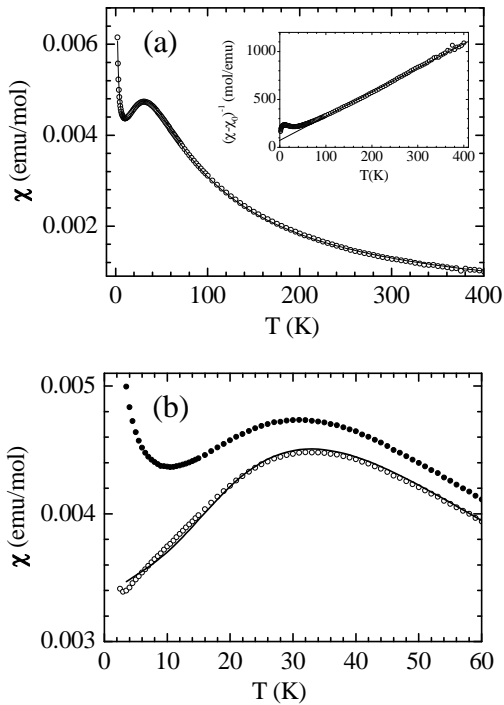}
\caption{The upper panel (a) shows the magnetic susceptibility versus temperature data in the temperature range 2 - 400 K with the applied field ($H$ = 10 kOe) parallel to the [201] direction. The solid line through the data points is a global fit of the data using a Johnston term, a temperature independent term, and a Curie term (as discussed in the text). The inset in the upper panel shows the inverse susceptibility versus temperature data with the temperature independent term ($\chi_0$) removed. The solid line in the inset shows a linear fit of the data in the temperature range 250 - 400 K. The lower panel (b) shows the low temperature (2-60 K) behavior of the susceptibility with  H = 10 kOe, applied parallel to  the  [201] direction. The data points with filled circles show the total susceptibility data, whereas the data points with open circles indicate susceptibility after the  Curie contribution ($\chi_{add}$) and the temperature independent positive term ($\chi_0$) have been  subtracted. The solid line through the open circles shows a Johnston model fit associated to the spin chain contribution ($\chi_{chain}$).}
\label{fig2}
\end{figure}

\section{Results}
\subsection{Susceptibility}
Figure 2 shows the magnetic susceptibility ($\chi = M/H$) versus temperature data of Sr$_3$CuPtO$_6$ in the temperature range 2-400 K with the applied magnetic field (10 kOe) along the [201] direction. The low temperature $\chi(T)$ data show a rise with decreasing temperature below 10 K, similar to that observed in a previous investigation~\cite{claridge} on this compound. Such a feature has also been observed  in a number of  other  spin chain compounds~\cite{matsuda,kiryukhin} and it is generally thought to arise from the presence of broken chains and/or paramagnetic impurities; both effects lead to a Curie like term.~\cite{liu} At intermediate temperatures, the susceptibility versus temperature  data shows a broad peak around 32 K, which is typical of  short range magnetic order in a spin chain system. The high temperature data (250-400 K) shows  Curie-Weiss  behavior along with a temperature independent positive term $\chi_0$. 
\par
For a quantitative analysis of the susceptibility versus temperature data, we express $\chi$ as a sum of three terms as in reference~\cite{motoyama} with 
\begin{equation}
\label{chiterms}
\chi = \chi_{add} + \chi_0 + \chi_{chain}
\end{equation}

The term $\chi_{add}$ = $c_{add}/(T - \theta_{add})$ is a Curie-like term coming from broken chains or paramagnetic  impurities in the system. The term $\chi_0$  is a combination of the  van Vleck paramagnetic term and the diamagnetic contributions from the atomic cores. $\chi_{chain}$ is the contribution coming from the 1D magnetic correlations among  the spins. This can be approximated using the Johnston model (equation (50) of reference~\cite{johnston}) for the $S = \frac{1}{2}$ uniform linear chain system where we write $\chi_{chain}(T)$ = $\chi\{J, g\}(T)$ where $J$  and $g$ (Land\'e $g$ factor) are  the free parameters. With these considerations, we have fitted our $\chi(T)$ data to equation~\ref{chiterms}~in the temperature range 2-400 K (solid line in Figure 2). The good quality of the fit over the full temperature range indicates that the Johnston model can be used to describe  the susceptibility behavior of Sr$_3$CuPtO$_6$. The parameters obtained from the fitting are: $J$ = 25.7(1) K, $g$ = 2.050(3), $\chi_0$ = 8.1 (7)$\times$10$^{-5}$ emu/mol, $c_{add}$ =0.0055(1) emu-K/mol and $\theta_{add}$ = -0.14 (5) K. Similar fitting for $T >$ 13 K ($\sim \frac{1}{2}J$), using the Bonner-Fisher model for $\chi_{spin}$, produces  a value of $J$ = 25.3(1) K, which compares well with the result obtained using the Johnston model. The value of $J$  is also  in good agreement with  the value (26.1 K) estimated by Nguyen {\it et al.}~\cite{science} using the Bonner-Fisher model. 

\par
The Curie constant associated with $\chi_{add}$ corresponds to a moment of 1.73 $\mu_B$ residing on  $\sim$ 1.5\% of the Cu atoms present in the sample. The value of $\chi_{add}$ remains practically unchanged if one sets $\theta_{add}$ = 0. This concentration of free Cu$^{2+}$ ions is comparable with the values previously reported in some other Cu$^{2+}$ based spin chain compounds~\cite{matsuda, kiryukhin}.

\par
At high temperatures (250-400 K), the 1/($\chi(T) - \chi_0$) versus $T$ plot  (see inset of the upper panel) is found to be linear with the value of $\chi_0$ obtained from the global fitting (2-400 K) of our data to equation~\ref{chiterms}. This is expected, because at high temperatures, $\chi_{add}$ becomes negligible and $\chi_{spin}$ reduces to a Curie-Weiss term. Additional confidence in  the reliability of this $\chi_0$ value comes from fitting the high temperature $\chi(T) - \chi_0$ data to a simple Curie-Weiss expression, where one finds an effective moment $p_{eff}$ = 1.77(2) $\mu_B$/Cu$^{2+}$, which is close to the theoretical value (1.73 $\mu_B$), and a paramagnetic Curie temperature $\theta_p$ = -26 ($\pm$ 2) K that compares well with the $J$ value as expected for an $S$ = $\frac{1}{2}$ spin chain system.

\par
The lower panel of figure 2 is an expanded view of the susceptibility data below 60 K, showing the total $\chi(T)$ (filled circles), the $\chi_{chain}$ data (open circles) and a fit using the  Johnston formula (solid line). It should be  noted that even on this expanded scale,  the overall agreement  of the $\chi_{chain}$ data to the Johnston fitting is found to be good. However, below  $\sim$ 20 K  a discrepancy between the experimental data and the fitted line appears. In fact, the change in the slope expected  in the Johnston model around $T \sim$ 0.4$J$ K (about 10 K in the present case), is not present in our data. Given the fact that Sr$_3$CuPtO$_6$ has an unconventional chain structure, some departures in the magnetic behavior from the theoretical model are not unexpected. Nevertheless, it must also be borne in  mind  that in the temperature range considered here, the $\chi_{chain}$ data is very sensitive to the subtraction of $\chi_{add}$, which may  introduce some uncertainty in the resulting  $\chi_{chain}$ values.

\par
In order to better understand the low temperature behavior of Sr$_3$CuPtO$_6$, it is useful to investigate the temperature dependence of the heat capacity of this system, which is expected to be much less affected by  additional contributions from broken chains or paramagnetic impurities.
 
\begin{figure}[t]
\centering
\includegraphics[width = 7.5 cm]{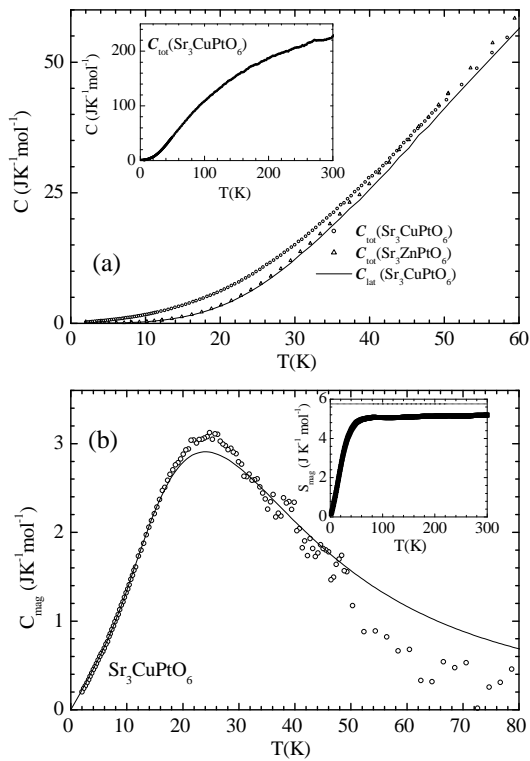}
\caption{The upper panel (a)  shows the heat capacity versus temperature data of Sr$_3$CuPtO$_6$ in the temperature range 1.9 - 60 K.  The heat capacity data of the nonmagnetic isostructural compound Sr$_3$ZnPtO$_6$ is also shown in the figure along with the lattice contribution of the heat capacity  of Sr$_3$CuPtO$_6$ as calculated (see text for details) from the heat capacity data of Sr$_3$ZnPtO$_6$. The inset in the upper panel shows a full  view (1.9 - 300 K) of the heat capacity versus temperature data of Sr$_3$CuPtO$_6$. The lower panel (b) shows the magnetic contribution to the heat capacity ($C_{mag}$) of Sr$_3$CuPtO$_6$ as a function of temperature along with the curve predicted by the Johnston model (solid line) for $J$ = 25.5 K. The inset shows the magnetic entropy versus temperature data for Sr$_3$CuPtO$_6$, and the dotted line in the inset denotes the theoretical value of the entropy ($Rln2$) for an $S$ = $\frac{1}{2}$ system.}
\label{fig3}
\end{figure}   

\subsection{Heat capacity} 
The total heat capacity of Sr$_3$CuPtO$_6$ as a function of temperature (1.9-300 K) is shown in figure 3a (inset). This data contains no signature of long range magnetic ordering  in this temperature range.  Figure 3 also depicts the $C(T)$ of the nonmagnetic isostructural compounds Sr$_3$ZnPtO$_6$. At low temperatures (below about 50 K), $C_{tot}$(Sr$_3$CuPtO$_6$) is larger in value than $C_{tot}$(Sr$_3$ZnPtO$_6$). This is clearly due to an additional  contribution to the heat capacity of magnetic origin for the Cu compound. It is expected that the magnetic contribution to  $C_{tot}$(Sr$_3$CuPtO$_6$) should tend to zero at high temperatures.
There is a difference between  $C_{tot}$(Sr$_3$ZnPtO$_6$) and  $C_{tot}$(Sr$_3$CuPtO$_6$) even at  300 K, which is due to the  mass difference of the Cu and the Zn atoms. In order to calculate the lattice contribution, $C_{lat}$, of Sr$_3$CuPtO$_6$ from the heat capacity data of Sr$_3$ZnPtO$_6$, a careful mass correction to the data is required.  In the present case we have used a phenomenological scaling of the Debye temperature ($\Theta_D$) with two  scaling factors (see appendix) to obtain the $C_{lat}$ from the heat capacity of the non-magnetic reference sample. It should  be noted that the scaling  for $C_{lat}$ has only a weak effect at low temperatures. $C_{lat}$(Sr$_{3}$CuPtO$_{6}$) is almost superimposed onto $C_{tot}$(Sr$_{3}$ZnPtO$_{6}$) below about 40 K. Since this is the region of interest for the analysis of the 1D magnetic features,  these  should not be significantly affected
by any imperfection in  the model used to derive $C_{lat}$(Sr$_{3}$CuPtO$_{6}$). 
 
\par
The lower panel of figure 3 shows the magnetic contribution to the heat capacity  of Sr$_{3}$CuPtO$_{6}$, $C_{mag}(T)$, that was obtained by subtracting the lattice contribution $C_{lat}(T)$ from the total heat capacity $C_{tot}(T)$. The main feature is a broad peak at low temperature, as expected in a 1D magnetic system~\cite{bonner}. In their theoretical investigation of $S = \frac{1}{2}$ Heisenberg antiferromagnetic chains, Johnston {\it et al.}~\cite{johnston} have proposed analytical expressions for the temperature dependence of the heat capacity. The solid line in figure 3 is the curve predicted by this model (equations 54a-c of reference~\cite{johnston}) for $J=25.5$ K. It is worth emphasizing that there are no free parameters other than $J$ in this model. On the scale of Figure 3, only curves  obtained using a narrow range ($\pm$ 0.5 K) of $J$  values around 25.5 K provide good overall agreement with the experimental data. The curve corresponding to $J$ = 25.5 K  gives the best fit for $T <\frac{1}{2}J$ and the quality of the agreement is quite satisfactory up to about 40 K. Above $T\sim 50$ K, the $C_{mag}(T)$ data clearly departs from the model, a behavior which can be ascribed to the  uncertainty in the estimation of $C_{lat}$ (for instance, at $T \sim$ 60 K, the expected  magnetic contribution is just 2\% of the measured total heat capacity). 
In the Bonner-Fisher model, the intra-chain coupling is related to the temperature $T_{max}$ corresponding to the  maximum in $C_{mag}(T)$ data as $T_{max} \simeq$ 0.962$J$. Considering $T_{max}$ in the range 24-25.5 K, one obtains $J$ value lying between 25 K and 26.5 K, which is in good agreement with the  value of $J$ given by the Johnston model above.

\par
The inset in the lower panel of figure 3 displays the temperature dependence of the magnetic entropy derived by $S_{mag}(T) = \int_0^T(\frac{C_{mag}(T')}{T'})dT'$.~\cite{entropy} One observes that the calculated entropy saturates slightly below the theoretical value, $R\ln 2$ = 5.76 J/mol-K, expected for  $S$ =1/2. This apparent {\it missing} entropy probably just reflects the uncertainty in the estimation of  $C_{lat}(T)$. 

\par
In order to get a more comprehensive view of the 1D magnetic  behavior of  Sr$_3$CuPtO$_6$, we have extended our heat capacity measurement down to 0.35 K. Figure 4 shows the $C_{mag}/T$ versus $T$  data below 11 K, while the inset in  figure 4 shows the temperature dependence of the $C_{mag}$ data below 6 K. Around 1.5 K, $C_{mag}$ contains a small  feature (which is more clearly visible in the $C_{mag}/T$ versus $T$ plot), and then  tends smoothly to zero as $T$ is  decreased further. This may be a signature of  three dimensional (3D) long range  magnetic ordering or of a spin-Peierls-like  transition. However, the occurrence of such a transition appears to be in contrast with the work of Irons {\it et al.}~\cite{irons}, who reported the absence of any anomaly in the a.c. susceptibility measurements down to 0.27 K on a polycrystalline sample of Sr$_3$CuPtO$_6$. In order to understand the significance of this feature, further investigations based on neutron scattering measurements are required. For the present analysis, we decided to restrict ourselves to the temperature range $T \geq $ 2 K, for which $C_{mag}$ is presumably related to  the short range intra-chain coupling only.

\par
In case of  an AFM spin chain system like Sr$_3$CuPtO$_6$, $C_{mag}$ is expected to show a linear $T $ dependence, $C_{mag} = \gamma_{mag}T$, at low temperatures (approximately, $T < J/5$  which is below 5 K in our case).  Our $C_{mag}(T)$ data in figure 4 (inset) shows a quasi-linear behavior in the temperature range 2-6 K, but clearly extrapolates to a finite $T$ intercept of about 0.5 K at $C_{mag}$ = 0. Such a behavior suggests the possibility of a small gap of $\Delta \sim$ 0.5 K in the spin excitation spectrum. Although, a spin gap should not exist in case of a  uniform $S = \frac{1}{2}$ spin chain system, the possibility of a gap cannot be ruled out in the case of Sr$_3$CuPtO$_6$, with its unconventional spin chain structure.

\par
In figure 4, we have shown $C_{mag}/T$ behavior predicted by the Johnston model (down to $T$ = 1 K) with $J$ = 25.5 K along with the experimental data. While the Johnston model well represents the experimental data down to 5 K, a  deviation is visible in the low temperature data. The experimental $C_{mag}/T$ contains a clear downward curvature  below 5 K, which is absent in the Johnston model that shows a rather constant $C_{mag}(T)/T$ behavior consistent with the expected linear $T$ dependence of the $C_{mag}$ data in this temperature range (1-5 K). This departure from the Johnston model behavior around 5 K  cannot be simply attributed to the small feature observed around 1.5 K.  If a spin gap is responsible for such a departure, to a first approximation, it can be accounted by a phenomenological relation $C_{mag}^{\Delta}/T = \gamma_{mag}^{\Delta} exp(-\Delta/T)$ at low temperatures. The dotted  line in figure 4 shows the fitting of $C_{mag}^{\Delta}/T$ to the data in the temperature range 2-5 K along with an extrapolation slightly below and above 2 K and 5 K respectively. The parameters estimated by the fitting  are found to be $\gamma_{mag}^{\Delta}$ = 0.128 JK$^{-2}$mol$^{-1}$ and $\Delta$ = 0.64 K. This phenomenological spin gap model can account for the downward curvature in the $C_{mag}(T)/T$ data below 5 K.  According to the Johnston model, for a $S = \frac{1}{2}$  AFM spin chain, the coefficient of the linear term ($\gamma_{mag}$) in $C_{mag}$ is related to the intra-chain coupling parameter as $\gamma_{mag} = R/3J$, where $R$ is the universal gas constant. Although, strictly speaking,  this relation is valid for an ideal gap-less spin chain system, we have used the relation in the present context to  estimate the value of $J$. The value of $J$, obtained  by inserting  the value of $\gamma_{mag}^{\Delta}$ in the above relation, is found to be 22.2 $\pm$ 0.5 K, which is close to the values of $J$ obtained from other methods.

\begin{figure}[t]
\includegraphics[width = 7.5 cm]{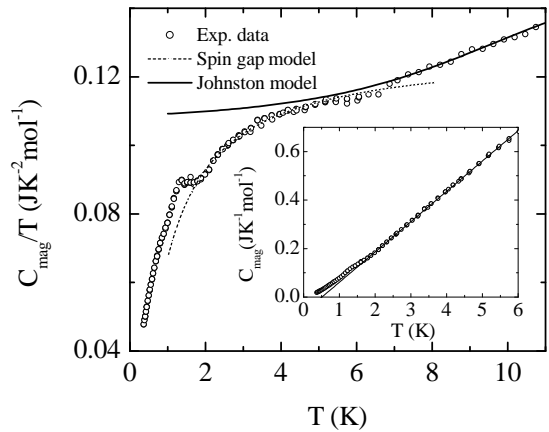}
\caption{The low temperature (down to 0.35 K) magnetic heat capacity versus temperature data of Sr$_3$CuPtO$_6$ along with the  curve (solid line) predicted by Johnston model (down to 1 K) for $J$ = 25.5 K. and the curve (dotted line) obtained by fitting  $C_{mag}^{\Delta} = \gamma_{mag}^{\Delta}.T. exp(-\Delta/T)$  to our data in the temperature range 2- 5 K and then extrapolated below and above 2 K and 5 K respectively. The inset shows the $C_{mag}$ versus $T$ data below 6 K. The line emphasizes the quasi-linear behavior of the data in the temperature range 2-6 K.}
\label{fig4}
\end{figure}  
   
\section{Discussion}
The present magnetic susceptibility and heat capacity  investigations on  single crystals of  Sr$_3$CuPtO$_6$ show that there is a good overall agreement between our experimental data and the  theoretical predictions for an  $S$ = $\frac{1}{2}$ Heisenberg AFM spin chain system. The values of the characteristic coupling parameter $J$, obtained from different  measurements, are reasonably consistent. 

\par
Such  good agreement with standard spin chain models is  rather surprising for a compound whose one-dimensional magnetic character was recently questioned~\cite{claridge, irons}. The doubts raised  were based upon the  large value of the inter-chain coupling parameter ($J'$), derived from an  analysis of the magnetic susceptibility, leading to a ratio $J/J' \sim$ 3. Our susceptibility analysis discussed so far in the text has been performed without considering any inter-chain coupling parameter (i.e., $J'$ = 0). In order to  get an idea of the inter-chain coupling parameter, we have reanalyzed our susceptibility data with  the relation proposed by Hatfield~\cite{hatfield} and  previously used by Claridge {\it et al.}~\cite{claridge}. We found that this mean field approach can improve the quality of the fitting with the addition of an  AFM inter-chain coupling parameter. The  best fit is obtained  for  $J$ = 25.1 ($\pm$ 0.1) K and $J'$ = 4.3 ($\pm$ 0.2) K. The $J'$ value is slightly less but still comparable with the value 7.3 K reported by Claridge {\it et al.}~\cite{claridge}. However, one should be careful in interpreting the value of $J'$ from this kind of analysis. For low dimensional magnetic systems (with small and weakly $T$-dependent $\chi$), the introduction of  an additional fitting parameter, $J'$,  can often lead to unphysical results. In the case of Sr$_3$CuPtO$_6$, such a large value of $J'$ appears to be inconsistent with the global magnetic behavior, particularly the absence of long range magnetic order down to 2 K. A finite value of the inter-chain coupling should eventually lead to a three dimensional  ordering at a temperature $T_N$ and theoretical models have established relationships between $T_N$ and the coupling parameters ($J$ and $J'$) for the weakly coupled linear chain systems. For $S$ = $\frac{1}{2}$ AFM  Heisenberg chains, the model proposed by Oguchi~\cite{oguchi}, based on a Green's function method, has been used extensively ~\cite{dejongh}.  Recently,  Schulz~\cite{schulz} has proposed another model relating $T_N$, $J$ and J$'$ for weakly coupled $S$ = $\frac{1}{2}$ chains, which is  based on a mean-field approximation for the inter-chain coupling and exact results for the resulting effective 1D problem.  Both these models deal with a square lattice of the spin chains and as outlined by Lines {\it et al.}~\cite{lines}, corrections are required to apply these models to  systems with a hexagonal arrangement of  spin chains, as is the case for Sr$_3$CuPtO$_6$.  With $J$ = 25.5 K, and $T_N \leq$ 2 K (since, there is no  3D order at least down to 2 K), both the Oguchi and the  Schulz models predict an  upper limit for  the inter-chain coupling constant $J'$ of 0.2 K. Therefore, it is clear that the large value of $J'$ derived previously, by fitting the Hatfield model to the  susceptibility data, may not be reliable. With $J' \leq$ 0.2 K, one obtains the {\it lower boundary} of the ratio of the intra-chain and inter-chain coupling parameters to be approximately 130 ($J/J' \geq$ 130) and this ratio is more consistent with the 1D magnetic character of Sr$_3$CuPtO$_6$.   

\par
Now the question is how to reconcile such a large $J/J'$ ratio with the fact that the distance between two neighboring Cu$^{2+}$ ions is approximately the same along and perpendicular to the chain. It is well known that the strength of the exchange coupling not only depends upon the distance, but also on the nature of the exchange pathway. From the viewpoint of the mediation of the exchange interaction between Cu$^{2+}$ ions, the situation is very different parallel and perpendicular to the chains. Although the Pt$^{4+}$ ions located in between the Cu$^{2+}$ ions along the chains are in the low spin state ($S$ = 0), they possess unfilled 3$d$ orbitals and can thus participate in the exchange interaction. A similar mechanism cannot be expected for the Sr$^{2+}$ ions located in between the CuO$_6$-PtO$_6$ chains. Such a difference in the exchange pathway can qualitatively explain the large $J/J'$ ratio in Sr$_3$CuPtO$_6$.

\par
Finally, let us consider  the low temperature (0.35-6 K) heat capacity data for Sr$_3$CuPtO$_6$, which revealed some deviation from the $S = \frac{1}{2}$ 1D AFM  Heisenberg model. Although, they are  also visible in $\chi_{chain}$ data (figure 3), such deviations can be studied more precisely in $C_{mag}$ data. The latter data led us to propose the  existence  of a small energy gap in the spin excitation spectra. A uniform half-odd integral spin chain system is supposed to be gapless~\cite{lieb}. One simple explanation for the origin of a small gap is a weak modulation in the strength of the exchange interaction along the chain~\cite{bonner2}. The zigzag nature of the spin chain in Sr$_3$CuPtO$_6$ may result in such a modulation.~\cite{carlin}.

\par
In conclusion, we have performed a detailed investigation of the magnetic susceptibility and the heat capacity behavior of the  compound Sr$_3$CuPtO$_6$. The use of a nonmagnetic iso-structural compound (Sr$_3$ZnPtO$_6$) has  enabled us to extract the magnetic contribution to the  heat capacity. Both the magnetic susceptibility and the heat capacity were found to be  consistent with the Johnston and Bonner-Fisher models for an $S = \frac{1}{2}$ AFM spin chain, with similar values of the intra-chain coupling parameters ($J \sim$ 25.5 K). Based on the fact that there is no long range magnetic order observed in this system, at least down to 2 K, the {\it lower limit} of the ratio of the intra-chain and inter-chain coupling parameters is expected to be $\sim$ 130. These observations, in contrast to the previous reports~\cite{claridge, irons}, clearly identify Sr$_3$CuPtO$_6$ as a $S$ = $\frac{1}{2}$ spin chain compound with 1D magnetic character. Deviations from the 1D uniform $S =  \frac{1}{2}$ Heisenberg models were observed in the low temperature regime, which indicate the possible existence of a gap in the spin excitation spectum.

\par
We acknowledge the financial support of the EPSRC (UK) for this project.

\appendix*
\section{Calculation of $C_{lat}$}
According to the Debye model of lattice heat capacity,
\begin{equation}
\centering
C_{lat}(T)=3pRD[\theta _{D}(T)/T]
\label{clat}
\end{equation}
where $p$ is the number of atoms per molecule, $R$ is the gas constant, $\theta _{D}(T)$ is the  Debye temperature, and $D(x)=(3/x^{3})\int_{0}^{x}[z^{4}e^{z}/(1-e^{z})^{2}]dz$ is the Debye function. The problem of finding $C_{lat}$ for a  magnetic sample reduces to the problem of  deriving the Debye temperature of the magnetic sample, $\theta _{D}^{sam}(T)$ from that of a non-magnetic isostructural reference compound, $\theta _{D}^{ref}(T)$, which can be directly derived from the total heat capacity $C_{tot}(T)$ of the reference sample by inverting equation~\ref{clat}. A one parameter scaling of the Debye temperature ($\theta _{D}^{sam}(T)$ =$r\theta _{D}^{ref}(T)$)(where $r$ is a constant) is found to be inadequate as there is no $r$ value leading to a $C_{lat}$(Sr$_3$CuPtO$_6$) which  tends to 
$C_{tot}$(Sr$_3$CuPtO$_6$) at 300 K, while lying below it for  $T <$ 300 K. The former requirement is based on the fact that at 300 K, the theoretical magnetic contribution to the heat capacity for $S = \frac{1}{2}$ Heisenberg chain with $J \sim$ 25 K, is expected to be only 0.02\% of the total heat capacity of Sr$_3$CuPtO$_6$. The failure of this one parameter scaling led us to use an extension of this approach with two phenomenological factors, corresponding to the highest ($T_H$) and lowest ($T_L$) temperatures of our data. We can express $\theta _{D}^{ref}(T)=\theta _{D}^{ref}(T_L)+[\theta _{D}^{ref}(T_H)-\theta _{D}^{ref}(T_L)]\;f(T)$, where $f(T)$ is a function such as $f(T_L)=0$ and $f(T_H)=1$. Owing to the structural equivalence of the magnetic and the nonmagnetic reference compounds, we assumed that $\theta _{D}^{sam}(T)$ can be approximated by a similar expression, keeping the same function $f(T)$. The $\theta _{D}^{sam}$ values at $T_L$ and $T_H$ are related to their counterparts in the reference sample by two adjustable parameters: $\theta _{D}^{sam}(T_L)=r_{1}\,\theta _{D}^{ref}(T_L)$, and $\theta _{D}^{sam}(T_H)=r_{2}\,\theta _{D}^{ref}(T_H)$. The parameter $r_{2}$ was determined from the requirement that $C_{mag}$ should vanish at $T_H$, resulting  in a $C_{lat}$ that  tends to $C_{tot}$ at $T_H$. For the present case of Sr$_3$CuPtO$_6$ with Sr$_3$ZnPtO$_6$ as the reference sample, the single parameter correction factor ($r$)(which is valid in the low $T$ region {\it i.e.,} $T \ll \Theta_D$ only) estimated from the difference in the mass of the atoms is almost equal to 1 ~\cite{bouvier}. Therefore, the parameter $r_{1}$ was chosen as the smallest value for which $C_{lat}$ does not cross $C_{tot}$ over the whole $T$-range. The estimated values of the parameters are:  $r_{1}=1.025$ and $r_{2}=1.050$ (with $T_L$ = 2 K and $T_H$ = 300 K). Figure 3 shows the corresponding lattice contribution of Sr$_{3}$CuPtO$_{6}$ that was obtained by including $\theta _{D}^{sam}(T)$ in the Debye formula.




\begin{thebibliography}{99}
\bibitem{stitzer} K. E. Stitzer, and H.-C. zur Loye, Current Opinion in Solid State and Material Science {\bf 5}, 535 (2001).

\bibitem{nguyen} T. N. Nguyen, D. M. Giaquinta, and H.-C. zur Loye, Chem. Mater. {\bf 6}, 1642 (1994).

\bibitem{claridge} J. Claridge, R. C. Layland, W. Hampton Henley, and H.-C. zur Loye, Chem. Mater. {\bf 11}, 1376 (1999).

\bibitem{science} T. N. Nguyen, P. A. Lee, and H.-C. zur Loye, Science {\bf 271}, 489 (1996).

\bibitem{hodeau} J. L. Hodeau, H. Y. Tu, P. Bordet, T. Fournier, P. Strobel, and M. Marezio, Acta Cryst. B {\bf 48}, 1 (1992).

\bibitem{hatfield} W. E. Hatfield, J. Appl. Phys. {\bf 52}, 1985 (1981).

\bibitem{irons} S. H. Irons, T. D. Sangrey, K. M. Beauchamp, M. D. Smith and H. C. zur Loye, Phys. Rev. B {\bf 61}, 11594 (2000).

\bibitem{niazi} A. Niazi, E. V. Sampathkumaran, P. L. Paulose, D. Eckert, A. Handstein, and  K.-H. M\"uller Phys. Rev. B {\bf 65}, 064418 (2002).

\bibitem{rayprol} S. Rayaprol, Kausik Sengupta, and E. V. Sampathkumaran, Phys. Rev. B {\bf 67}, 180404(R) (2003).

\bibitem{kausik}Kausik Sengupta, S. Rayaprol, Kartik K. Iyer, and E. V. Sampathkumaran, Phys. Rev. B {\bf 68} 012411 (2003).   

\bibitem{hardy} V. Hardy, S. Lambert, M. R. Lees, and D. McK. Paul, Phys. Rev. B {\bf 68}, 014424 (2003).

\bibitem{bonner} J. C. Bonner and M. E. Fisher, Phys. Rev. {\bf 135}, A640 (1964).

\bibitem{johnston} D. C. Johnston, R. K. Kremer, M. Troyer, X. Wang, A. Kl\"umper, S. L. Bud'ko, A. F. Panchula, and P. C. Canfield, Phys. Rev. B {\bf 61}, 9558 (2000). 

\bibitem{klumper} A. Kl\"umper, and D. C. Johnston, Phys. Rev. Lett. {\bf 84}, 4701 (2000).

\bibitem{eat} S. Eggert, I. Affleck, and M. Takahashi, Phys. Rev. Lett. {\bf 73}, 332 (1994).

\bibitem{lukyanov} S. Lukyanov, Nucl. Phys. B {\bf 522}, 533 (1998).

\bibitem{christina} C. L.-\"Onnerud, M. Sigrist, and H.-C. zur Loye, J. Solid State Chemistry {\bf 127}, 25 (1996).

\bibitem{matsuda} M. Matsuda and K. Katsumata, Phys. Rev. B {\bf 53}, 12201 (1996).

\bibitem{kiryukhin} V. Kiryukhin, Y. J. Kim, K. J. Thomas, F. C. Chou, R. W. Erwin, Q. Huang, M. A. Kastner, and R. J. Birgeneau, Phys. Rev. B {\bf 63}, 144418 (2001).

\bibitem{liu} Y. Liu, J. E. Drumheller, and R. D. Willett, Phys. Rev. B {\bf 52}, 15327 (1995).

\bibitem{motoyama} N. Motoyama, H. Eisaki and S. Uchida, Phys. Rev. Lett. {\bf 76}, 3212 (1996).

\bibitem{entropy} In order to calculate $S_{mag}$, we have used our $C_{mag}(T)$ data in the temperature range 0.35 to 300 K and extrapolated down to 0 K.

\bibitem{oguchi} T. Oguchi, Phys. Rev. {\bf 133}, A1098 (1964).

\bibitem{dejongh} L. J. De Jongh  and A. R. Miedema, Adv. Phys. {\bf 23}, 1 (1974).

\bibitem{schulz} H. J. Schulz, Phys. Rev. Lett. {\bf 77}, 2790 (1996).

\bibitem{lines} M. E. Lines, and M. Eibsch\"utz,  Phys. Rev. B {\bf 11}, 4583 (1975).

\bibitem{lieb} E. Lieb, T. D. Schulz and D. C. Mattis, Ann. Phys. {\bf 16}, 407 (1961).

\bibitem{bonner2} J. C. Bonner, and H. W. J. Bl\"ote, Phys. Rev. B {\bf 25}, 6959 (1982).

\bibitem{carlin} R. L. Carlin, {\it Magnetochemistry} (Springer-Verlag, Berlin, 1986), p. 75.

\bibitem{bouvier}M. Bouvier, P. Lethuillier, and D. Schmitt, Phys. Rev. B, {\bf 43}, 13137 (1991). 

\end{thebibliography}
\end{document}